# Rotational Dynamics of ATP Synthase: Mechanical Constraints and Energy Dissipative Channels

Islam K. Matar,[(a,b)] Peyman Fahimi,[(c)*] Chérif F. Matta[(a,b)*]

(a) Department of Chemistry and Physics, Mount Saint Vincent University, Halifax, NS B3M 2J6 Canada.
(b) Department of Chemistry, Saint Mary's University, Halifax, NS B3H 3C3 Canada.
(c) Department of Mathematics and Statistics, Dalhousie University, Halifax, NS B3H 4R2 Canada.
* E-mail: fahimi@dal.ca, cherif.matta@msvu.ca

## Abstract

The proton motive force (PMF) across the inner mitochondrial membrane delivers approximately 0.2 eV of energy per proton, powering the $F_oF_1$-ATP synthase molecular motor. Here, we provide a detailed accounting of how this energy is utilized: Approximately 75–83% is transduced into the chemical free energy of ATP synthesis, while the remaining 17–25% is dissipated through internal friction, viscous drag, proton leakage, electroviscous effects, elastic deformations, and information-theoretic costs. Each dissipation channel is quantitatively evaluated, revealing that internal friction in the $F_1$ motor is the dominant loss mechanism. In this work, we did not account for the energy supplied/injected due to the intrinsic electrostatic potential of the enzyme itself. In addition to this energy bookkeeping, we also examine the quantum mechanical constraints on the $F_o$ unit's rotation. We find that, as can be expected, the energy spacing between quantized rotational states is several orders of magnitude smaller than thermal energies at physiological temperature, and that the tunneling probability through rotational barriers practically totally non-existent. Furthermore, the biological rotation speed (~100–650 revolutions per second (rps)) is between one and three orders of magnitude below the quantum limit implied by quantization of angular momentum of the c-ring (which would have been *ca.* 13,000 to 62,000 rps (depending on the size of the c-ring (17 to 8 subunits, respectively)) in the first rotational energy level of the c-ring). Nevertheless, experimental estimates of the rotation rates in *isolated* c-ring suggest rates in the vicinity of 43,000 rps, right within our theoretical quantum estimates. However, ATP synthase *as a whole* operates firmly within the classical regime, despite its nanoscale dimensions, and highlight its evolutionary optimization for robust and efficient energy conversion at the quantum-classical boundary. This is the result of the rotatory coupling between the $F_o$ and the much slower $F_1$ unit. ATP synthase's purely classical behavior showcases a remarkable evolutionary optimization of one of life's most essential rotary motor engineered so as to thrive far from the quantum limit, securing its function against the uncertainties of the quantum world. As Schrödinger stated in *What is Life?* (1944): "*The submicroscopic world is full of fluctuations. But in large aggregates of atoms, the law of large numbers ensures that these fluctuations become negligible*" a prediction directly confirmed here in the context of the rotational stability of the $F_o$ unit.

## 1. Introduction

Recent advances in quantum biology have demonstrated that quantum effects can influence biological processes such as photosynthesis, enzyme dynamics, spontaneous and induced mutations, and molecular evolutions.[1-26] However, the role of quantum mechanical constraints on large, dynamic biological machines - such as ATP synthase - remains largely unexplored. An important example of studies, few and far between, focused on quantum mechanical effects related to the operation of ATP synthase is the recent report of Uzoigwe.[27] This author argues that nuclear quantum effects (NQEs) related to protons are essential for explaining the efficiency of chemiosmosis and of ATP synthesis by oxidative phosphorylation. Uzoigwe shows that proton delocalization and quantum coherence enable proton transport across ATP synthase even under minimal proton gradients. "*Quantum chemiosmosis*" is proposed as a hybrid model integrating classical and quantum physics to explain the efficiency of mitochondrial ATP production. We will pick up this overarching conceptual thread laid out by Uzoigwe, that is, further explore some of the quantum mechanical aspects as they relate to the operation of ATP synthase.

While ATP synthase has been extensively characterized classically, to the best of the authors' knowledge there appears to be no study to date that systematically assesses whether quantum limits on rotation, such as angular momentum quantization, energy level discretization, or tunneling phenomena, impose fundamental restrictions on its function. While ATP synthase, with its 30,000–40,000 non-hydrogen atoms can, in principle, be treated as a quantum object, quantum effects in such a large and complex molecule are expected to be negligible compared to classical ones. One can expect negligible quantum behaviour of this enzyme given the extremely small de Broglie wavelength due to the molecule's large mass ($\lambda = h/p$, where $h$ is Planck's constant and $p$ the momentum), the vast number of internal degrees of freedom (vibrations, rotations, conformational transitions) which would lead to rapid internal decoherence, and the enzyme's interaction with its environment (e.g., membrane and solvent, the thermal random Brownian motion at physiological temperatures, etc.). The largest molecule in which matter-wave interference has been experimentally observed contains only ~ 2,000 atoms and requires highly controlled vacuum and temperature conditions.[28] That being said, we wish to elucidate *on what side of the "fence" does ATP synthase fall: The quantum or the classical?* Hence, we include back-of-the-envelope calculations, as no such analysis appears to exist in the literature. Understanding the operational limits of this enzyme is crucial not only for deepening our knowledge of molecular biophysics, but also for guiding the design of synthetic nanomotors[29-31] and biomimetic devices.[32-35] In addition to its relevance in biophysics and nanotechnology, ATP synthase also holds significant biomedical importance. For instance, the differences between mycobacterial and human oxidative phosphorylation machinery, especially ATP synthase, have been reviewed recently by Matar *et al.*, offering insights for selective inhibition strategies that would inhibit bacterial ATP synthase leaving the human enzyme unaffected.[36] Given the emergence of quantum-enabled nanotechnology and the growing interest in quantum effects in biology, we seek to establish, with a quantitative analysis, whether biological rotors like ATP synthase operate within classical or quantum regimes.

## 2. The limits on the rate of rotation of the $F_o$ unit of ATP synthase due to quantization of angular momentum

As is well-known, the $F_o$ is a rotary motor embedded in the inner mitochondrial membrane (IMM), in the thylakoid membrane of chloroplasts, plasma membrane of bacteria, etc.[37-39] While the magnitudes used in the estimates outlined below relate primarily to *mitochondrial, chloroplast, and bacterial ATP synthases*, the same ideas can be extended and applied to different molecular nanomachines in biology or nanoscience.

To better understand the mechanistic basis of proton translocation in such systems, several groups have developed detailed computational models of ATP synthase function. Ivontsin *et al.* apply a quantum-mechanical model combined with stochastic methods to simulate proton transport in the $F_oF_1$-ATP synthase.[40] These authors show that proton movement depends critically on the electrostatic landscape shaped by charged centers (amino acids and water molecules arranged in series) inside the half-channels, with quantum tunneling facilitating proton hopping. Their Monte Carlo simulations calculated a mean proton transfer time of about 23 ms, highlighting key regions requiring water occupancy for efficient proton transport.[40] However, a rate of 23 ms per proton corresponds to approximately 44 protons per second, which is significantly lower than the experimentally observed rate of around 1000 protons translocating through ATP synthase per second.[41,42] Ivontsin *et al.*[40] reported mean proton transfer time from a hybrid approach: quantum mechanics is used to compute tunneling probabilities between binding sites, while time is reintroduced via semiclassical estimates of transfer rates. By simulating many stochastic pathways using Monte Carlo methods, they statistically extract an average dwell time. Thus, the timescale reflects semiclassical transport dynamics superimposed on a quantum-informed energy landscape.

Ivontsin *et al.* follow-up their previously mentioned work with a study in which they use a quantum-mechanical and stochastic approach to identify the key amino acid residues involved in proton transport through the inlet half-channel of $F_oF_1$-ATP synthase.[43] The authors identify His245 and Asp119 as critical residues that significantly affect the efficiency of proton transfer. The results of the Monte Carlo simulations by these researchers yield estimates of proton hopping times on the nanosecond timescale between two adjacent centers (amino acid or water molecule) depending on the specific amino acid residues involved in the transfer,[43] but the total time for a proton to traverse the entire $F_o$ inlet half-channel is in the millisecond range - as mentioned above, reflecting the cumulative effect of many such hops and the "residence times" between hops on the difference transmission nodes.[40]

According to the Boyer-Walker mechanism,[44-46] the c-subunit of the $F_o$ unit rotates as protons pass through it along the electrical and concentration gradients. Coulombic force in the channel between the a- and c-subunit (the interface between them) favors one direction of rotation and overwhelms Brownian fluctuations into this direction[47] (clockwise during ATP synthesis when seen along the long axis of the enzyme from the mitochondrial side).[48] Given this rotation of a nano-scale structure, we enquire, here, whether *angular momentum quantization* sets a limit on the rate of this rotation.[49]

The simplest textbook treatment of angular momentum is that of the rigid rotor. ATP synthase as a whole *is not* strictly a rigid rotor since it undergoes torsion through its axel.[50,51] Since we are only interested in semi-quantitative answers in this paper, and as a first approximation, if we assume that the rotation of the $F_o$ unit behaves like a nano-rigid rotor, then its rotational angular momentum must be quantized (even if we expect the energy level separations to be tiny (but how tiny?)). The quantum mechanical angular moment $L$ is expressed as:

$$L = \sqrt{l(l+1)}\hbar, \quad (1)$$

where $l$ is the angular momentum quantum number (integer or half-integer) and $\hbar = h/2\pi = 1.054571817 \times 10^{-34}$ J·s is the reduced Planck's constant. Meanwhile, the classical definition of angular momentum is:

$$L = I\omega, \quad (2)$$

where $I$ is the moment of inertia of the rotating part of the $F_o$ unit (the c-ring) and $\omega$ is the angular velocity.

Equating the angular momentum from these two expressions, and solving for $\omega$ we obtain:

$$\omega = \frac{\sqrt{l(l+1)}\hbar}{I}, \quad (3)$$

which upon substituting the smallest nonzero angular momentum $l = 1$ (since $l = 0$ corresponds to no rotational motion), we get:

$$\omega_{\min} = \frac{\sqrt{2}\hbar}{I}. \quad (4)$$

It is important to note that for full unit consistency in rotational dynamics, the moment of inertia $I$ carries hidden radians in its units (kg·m²/rad² rather than just kg·m²).[52] As discussed by Oberhofer, radians are not to be treated as merely dimensionless in physical equations but must be accounted for to preserve the correctness of unit analysis.[52] This ensures that Eq. (4) properly balances dimensionally, yielding angular velocity $\omega$ in units of rad·s$^{-1}$, as expected. [52]

## 3. Estimates for the moment of inertia (*I*) and the angular velocity $\omega$ of the F$_o$ unit

F$_o$ is the portion of ATP synthase embedded in the membrane and is dominated by a cylindrical subunit called the "c-ring" composed of a number of α-helical units that vary by organism, from 8 in humans,[53,54] from 13 to 15 in cyanobacteria,[55] and from 9 to 17 in heterotrophic bacteria.[56] The c-ring assumes a cylindrical geometry and is the element of structure within ATP synthase that undergoes the rotation that drives the axel to rotate inside the F$_1$ unit forcing the three catalytic active site to switch between the O "*open*", L "*loose*", and T "*tight*" configuration every full rotation. We can, hence, model this physical element as a rotating cylinder. The moment of inertia for a cylinder rotating about its axis is given by:

$$I = \frac{1}{2}MR^2 \qquad (5)$$

in which, for an 8-subunit c-ring in human ATP synthase,[57] the mass is approximately 71,528 Da ($\approx 1.19 \times 10^{-22}$ kg), while for 17 c-rings in the human pathogen *Burkholderia pseudomallei*, the mass is approximately 141,000 Da ($\approx 2.34 \times 10^{-22}$ kg).[58] These values can be calculated using the FASTA sequence of each c-subunit from any ATP synthase structure available on the RCSB website (such as the one reported by Zhang *et al.*,[57] PDB Entry: 8KI3),[59] combined with the molecular weight of amino acids, which can be obtained from a standard reference.[60] As for the radius of the c-ring, that ranges from 2.68 nm for 9-subunit c-rings to 3.6 nm for 15-subunit c-rings.[61] A simple calculation reveals that for each additional c-subunit, the radius increases by approximately 0.15 nm ($\frac{3.6-2.68}{15-9} \sim 0.15$). This gives an approximate radius of 2.53 nm for 8 c-rings and 3.9 nm for 17 c-rings. Inserting the lower and upper bound values of the mass and radius of c-rings results in $I \approx 3.81 \times 10^{-40}$ kg.m$^2$ to $I \approx 1.78 \times 10^{-39}$ kg.m$^2$. Inserting these values into Eqn. (4) we obtain $\omega_{min} \approx 8.34 \times 10^4$ rad.s$^{-1}$ to $\omega_{min} \approx 3.90 \times 10^5$ rad.s$^{-1}$, which upon division by $2\pi$ translates into approximately $1.33 \times 10^4$ rps in 17 c-rings bacterial ATP synthase to $6.21 \times 10^4$ rps in 8 c-rings human ATP synthase.

Our calculation shows that the rate of rotation of the F$_o$ unit of ATP synthase as an ideal quantized nanoscale rigid rotor ranges approximately between 13,000 and 62,000 rps. However, the actual rotation rates of ATP synthase are significantly slower and vary depending on the species and temperature. Typical values reported in the literature range from 100 to 150 rps.[39] Higher rates have also been observed, such as 280 rps for yeast mitochondrial F$_1$,[62] 300 rps for *E. coli* F$_o$F$_1$,[63] 310 rps for bovine mitochondrial F$_1$,[64] and a range from 230 rps at 25°C to 650 rps at 45°C for isolated intact thermophilic F$_o$F$_1$ from recombinant *E. coli*.[65] Extrapolating this temperature-dependent trend to organisms with higher optimal growth temperatures suggests a theoretical rotation speed of up to 1600 rps at 60 °C - though this remains to be experimentally confirmed.[65] It is important to remind the reader here that the rates of rotation of F$_1$ and of F$_o$ are coupled and approximately equal (F$_o$ unit rotation rate ≈ F$_1$ unit rotation rate) since they are linked mechanically via the central stalk (the γ-subunit). One full rotation (360°) generates 3 ATP molecules.

A study by Lill, Althoff, and Junge[66] measures a maximum proton translocation rate of approximately $6 \times 10^5$ protons per second through the *isolated* c-ring of chloroplast ATP synthase in a voltage difference of 100 mV. Assuming 14 protons per full rotation, this corresponds to about 43,000 rps, falling within the quantum mechanical frequency estimated in our analysis (13,000 to 62,000 rps). If this is the observed speed with 100 mV potential difference, the value under larger voltages (e.g., the usual 200 mV observed in mitochondria) can only be higher. However, the $F_1$ unit in the *complete* ATP synthase protein constitute the rate-limiting rotatory step bringing down the rotation speed to the observed typical range of 100-650 rps under physiological conditions.

The moment of inertia of $F_o$ is sufficiently large that the spacing of the rotational quantum levels are extremely close together. The rotational energy levels are given by:

$$E_l = \frac{\hbar^2}{2I} l(l+1), \qquad (6)$$

with $l = 0, 1, 2, 3, \ldots$ . Thus, the energy spacing between adjacent rotational levels (say, between $l = 0$ and $l = 1$) is:

$$\Delta E = E_1 - E_0 = \frac{\hbar^2}{I}, \qquad (7)$$

which using the value of $I$ obtained from Eqn (5) yields a range of $\Delta E \approx 2.89 \times 10^{-29}$ J $\approx 1.80 \times 10^{-10}$ eV and $\Delta E \approx 6.19 \times 10^{-30}$ J $\approx 3.86 \times 10^{-11}$ eV. Thermal energy ($k_B T$) at room temperature ($T \approx 300$ K) is about 0.025 eV,[67] and thus, thermal fluctuations completely blur these energy levels and the motor rotates smoothly. By that we mean that the transitions between rotational levels for ATP synthase appear continuous and not quantized for all practical purpose. The angular momentum can increase by arbitrarily small chucks concealing the extremely minute quantized jumps.

Let us now also consider quantum mechanical tunneling using the rotational WKB (Wentzel–Kramers–Brillouin) approximation.[68,69] The tunneling probability is given by:

$$P \sim e^{-2S/\hbar}, \qquad (8)$$

where the action integral (S), *i.e.*, the integral of the difference between kinetic energy and potential energy over time as the system evolves, for an approximately constant potential (a flat barrier) is given by:

$$S \approx \Delta\theta \sqrt{2I(U_0 - E_{\text{rot}})}, \qquad (9)$$

where $\Delta\theta$ is the angular barrier width – *i.e.*, the angular width the system must tunnel across, the barrier height is given by $U_0$ and $E_{\text{rot}}$ is the total rotational kinetic energy of the $F_o$ unit. We model the resisting potential to rotation as coming from uniform friction (a single energy barrier), then we assume a flat barrier over a small angular distance $\Delta\theta \ll$ as a first approximation for tunneling in ATP synthase-like systems.

The barrier for the entry of a proton into the a-subunit has previously been estimated to range from 0.5 kJ/mol ($\approx$ 0.005 eV per molecule) in the ATP synthase of *Saccharomyces cerevisiae* to 3 kJ/mol ($\approx$ 0.031 eV per molecule) in *Yarrowia lipolytica*;[70] we take these values as $U_0$. Taking the potential difference across the two faces of the IMM to be ~ 200 mV, the total energy released per proton as it is translocated through ATP synthase is $E_{total} = qV \approx 3.204 \times 10^{-20}$ J $\approx 0.20$ eV/proton. Meanwhile, the amount of this energy converted to rotational kinetic energy must be *considerably smaller* (*vide infra*) than the proton motive force. This allows us to make the approximation that $E_{rot} \approx 0$. Finally, assuming a one c-subunit hop corresponds (in human ATP synthase with 8 c-subunits and in Bacterial ATP synthase with 17 c-subunits) to about:

$$\begin{cases} \Delta\theta_{c8} = \frac{2\pi}{8} \approx 0.785 \text{ radians} \\ \Delta\theta_{c17} = \frac{2\pi}{17} \approx 0.369 \text{ radians} \end{cases} \quad (10)$$

which upon substitution in Eq. (9) we obtain:

$$\frac{S}{\hbar} = \begin{cases} \Delta\theta_{c8}\sqrt{\frac{2I_{c8}U_{0-low}}{\hbar^2}} \approx 5.84 \times 10^3 \\ \Delta\theta_{c8}\sqrt{\frac{2I_{c8}U_{0-high}}{\hbar^2}} \approx 1.45 \times 10^4 \\ \Delta\theta_{c17}\sqrt{\frac{2I_{c17}U_{0-low}}{\hbar^2}} \approx 5.94 \times 10^3 \\ \Delta\theta_{c17}\sqrt{\frac{2I_{c17}U_{0-high}}{\hbar^2}} \approx 1.48 \times 10^4 \end{cases} \quad (11)$$

which gives a totally negligible tunneling probability of ~ exp(−14,500) to exp(−5,840), as expected all along.

We now return to the reason for the assumption that $E_{rot} \approx 0$. In our tunneling probability estimates, we set the rotational kinetic energy to zero to evaluate the most conservative scenario for quantum barrier penetration. This choice yields a strict lower bound on the tunneling probability, since any nonzero initial rotational energy would raise the system's total energy and lower the effective barrier height ($V_0 - E_{rot}$), thereby increasing the tunneling probability. Thus, our zero-energy assumption ensures that the calculated tunneling probability is a minimum value, reinforcing the conclusion that tunneling is negligible under physiological conditions.

The energy available per proton is ~ 0.20 eV is used by ATP synthase to rotate $F_o$ and to drive ATP synthesis. Because the ATP synthase operates in a highly overdamped regime,[71] the rotational kinetic energy of the c-ring is negligibly small compared to the energy delivered by the proton motive force per proton. Almost all of the input energy is immediately expended to overcome viscous, frictional, and conformational resistances rather than being converted as kinetic energy. The available potential energy is spent to overcome friction, do mechanical work, pay the energy cost of synthesizing ATP, and dissipate energy related to the action of ATP synthase as a reverse Maxwell demon. At 0.20 eV per proton, and about 8/3 protons (0.53 eV) to 17/3 protons (1.13 eV) needed to synthesize one ATP molecule, this exceeds the known free energy change of ATP synthesis ($\Delta G \sim$ 30-50 kJ/mol $\approx$ 0.3–0.5 eV per ATP) by a factor of the order of 100%. The part of $\Delta G$ which is *not* captured in ATP synthesis can be expended to overcome the viscous drag from the c-ring rotation through lipid bilayer membrane, the electroviscous drag due to the electrostatic interaction of the charged protein residue with the electric fields in which it is embedded, the internal friction inside the $F_1$ motor whereby overcoming molecular mechanical resistance is achieved during the conformational changes of $F_1$ to achieve catalysis, proton leaks (that is, protons crossing the bilayer itself through water wires and adenine nucleotide translocases, hence, bypassing ATP synthase through short-circuiting the pathway that is normally closed by the current returning to the matrix through ATP synthase),[72,73] thermal noise (Brownian motion) which are manifested as random torques opposing the general direction of the $F_o$ rotation, and finally, non-ideal transmission of torque between $F_o$ and $F_1$ (structural elasticity and slippage).[50] Electroviscosity[74-76] refers to the enhancement of the effective viscosity of a fluid due to an imposed electric field or to an ionic environment. In the case of ATP synthase *while embedded in the inner mitochondrial membrane*, electroviscosity that results from the electric double layer forming around the protein – that is, *primary* electroviscosity - is not expected to play a major role since the rotating part is surrounded by the hydrophilic tails of the phospholipid bilayer (save the two thin layers at the two surfaces of membrane which are highly charged). Meanwhile, the interactions of the (strong) electric field perpendicular to the membrane plane due to the proton imbalance can be a source of significant (*secondary*) electroviscosity.[77] Now let's provide estimates of all these dissipatory mechanisms.

## 4. Dissipative mechanisms associated with the $F_o$ rotation

### 4.1 Internal Friction Inside $F_1$

To estimate the energy cost due to internal conformational changes associated with torsional friction, and proceeding from an assumed "maximum" torque needed to rotate the $F_1$ unit of 40 pN.nm,[78,79] we get:

$$W_{\text{internal}} \approx 40 \text{ pN.nm} \times 2\pi/3 \approx 83.8 \text{ pN.nm} \approx 83.8 \times 10^{-21} \text{ J} \approx 0.5 \text{ eV}. \quad (12)$$

Thus, the internal friction $\approx$ 0.5 eV per ATP synthesized. Since the translocation of approximately 2.67 to 5.67 protons drive the synthesis of 1 ATP, then the energy dissipated by friction per proton should be around $\approx$ 0.09 eV to 0.19 eV, which is, hence, a major dissipative mechanism for chemiosmotic energy.

## 4.2 Viscous (and electroviscous) drag between $F_o$ and the inner mitochondrial membrane

The torque on a rotating rigid cylinder due to viscous drag in a Newtonian fluid, within the low Reynolds number regime, can be approximated by the expression:

$$\tau_{\text{viscous}} \sim 4\pi\eta L R^2 \omega, \tag{13}$$

where $\eta$ is the viscosity of the membrane (0.1 to 0.2 Pa.s for cell membrane at 37°C,[80] and ~ $0.8\times10^{-1}$ Pa.s for nematic polar liquid crystals at 35°C under a field strength of $10^6$ V/m.[81] For comparison, water viscosity is ~ $0.7\times10^{-3}$ Pa.s at 35°C. Here, L is the length of the c-ring, taken to be approximately equal to the average membrane thickness (~4 nm),[82] and $R$ is the radius of the rotor ($R_{c8}$ = 2.53 nm and $R_{c17}$ = 3.9 nm). The angular velocity is in the range $\omega = 2\pi \times 1,00 \approx 6.28 \times 10^2$ rad/s to $\omega = 2\pi \times 650 \approx 4.08 \times 10^3$ rad/s, corresponding to rotational speeds of 100 to 650 rps. The work per rotation is, thus:

$$W = \tau_{\text{viscous}} \times 2\pi \approx \begin{cases} 1.27 \times 10^{-22} \text{ J} \approx 0.0008 \text{ eV} \\ 3.92 \times 10^{-21} \text{ J} \approx 0.0244 \text{ eV} \end{cases}. \tag{14}$$

Thus, the viscous drag spans from 0.0008 eV to 0.0244 eV per full rotation. The contribution per proton depending on the number of c-subunits varies between $10^{-4}$ eV and $10^{-3}$ eV.

To the viscous drag we can also add electroviscous drag which usually a small correction (around 5–20%) to ordinary viscosity. The electroviscous drag is primarily determined by the surface electrostatic potential (zeta potential) of the c-ring, the electrokinetic radius (which accounts for the c-ring's physical radius plus the surrounding electric double-layer thickness) and the medium's permittivity.[83] Thus, electroviscous energy dissipation should be at most $\approx 1.6\times10^{-4} - 4.9\times10^{-3}$ eV per full rotation.

The extremely small contribution from viscous (electro)drag is consistent with arguments that suggest that evolution has fine-tuned the rotatory part of $F_o$ unit to be "well lubricated" with hydrophobic (oily) residues confronting one another to reduce wasteful energy dissipation through this channel. See discussion in Ref. [84].

## 4.3 Other energy dissipative processes

Brownian motion introduces random torques that cause small but unavoidable energy dissipation in rotary motors. In ATP synthase, Junge et al.[71] discuss the effects of thermal noise and viscous drag in opposing rotation from which it is subsumed that energy is primarily dissipated through internal friction. Johnson[85] emphasizes the inevitability of small energy losses ia random Brownian motion in molecular machines. Noji et al.[78], from direct observations of the $F_1$ rotation, indicate that although movement is quantized in steps, thermal agitation is sufficiently small compared to the work done per ATP synthesized suggesting Brownian-related dissipation must be an order of magnitude lower, *i.e.*, to the tune of ~0.01 eV. Thus, thermal noise in the form of random Brownian torques add probably a small additional dissipation which is ≈ 0.01 eV per proton (and hence 0.08 eV per rotation in humans).

A similar magnitude of dissipation is expected from structural elastic deformation in the stalk (γ-subunit) and at the substrate binding sites which could result, also, in a minor energy loss of perhaps ≈ 0.01–0.02 eV per proton (≈ 0.08–0.16 eV per rotation in humans). This is consistent with molecular simulations and models of torque generation and elastic power transmission in ATP synthase.[50,71]

Proton leak across mitochondrial membranes can dissipate some 20% of total proton flux,[72,73] representing a loss of ≈ 0.04 eV per proton (≈ 0.32 eV per rotation in humans and ≈ 0.68 eV per rotation in bacteria with 17 c-subunits). And finally, the operation of ATP synthase as a reverse Maxwell's Demon[86,87] and by Landauer's principle[88] requires a minimal energy dissipation per erased bit of information:[89,90]

$$E_{\text{bit}} = k_B T \ln 2, \quad (15)$$

leading, at room temperature, to $E_{\text{bit}} \approx 2.87 \times 10^{-21}$ J ≈ 0.018 eV (per translocated proton), i.e. ≈ 0.14 eV per rotation in humans. This is the bear minimum in a noiseless medium but the actual value is probably larger in the "noisy" (hot)[51,91-93] mitochondrial environment.[85,94,95]

## 4.4 Summary of dissipative processes

Table 1 and Fig. 1 provide a summary of the major contributors in the dissipation of energy. The energy delivered per proton moving across the inner mitochondrial membrane is ≈ 0.2 eV at the typical proton motive force (PMF) of around V ≈ 200 mV.[96] During the operation of ATP synthase, this energy is dissipated through multiple channels. Internal friction within the $F_1$ motor consumes the largest portion, approximately 0.09 eV to 0.19 eV (70% of proton energy on average), followed by proton leak (~0.04 eV, 20% on average), Maxwell Demon information-theoretic dissipation (~0.018 eV, 9%), structural elasticity and slippage (~0.015 eV, 7.5%), thermal noise (~0.01 eV, 5%), viscous drag from the membrane (~0.0005 eV, 0.3% on average), and electroviscous drag (~0.0001 eV, 0.05%). Depending on the biological conditions, the total dissipative losses combined is approximately 0.17 – 0.28 eV per proton. Of course, conservation of energy prohibits the sum of these dissipative processes to exceed the available energy in the first place (0.2 eV). The reason the total dissipation estimated here exceeds this threshold is that we simply added the *maximum possible losses across different dissipation pathways without adjusting for the actual energy available*. Clearly, some of the dissipative channels are alternative and, hence, are not strictly cumulative. Further, not all dissipation channels operate independently at their maximum values per proton. Thus, the simple sum of viscous drag, friction, leak, thermal noise, information dissipation etc. overcounts by necessity. Let us now estimate the actual total energy losses by the various channels of dissipations discussed above.

The dissipated energy is defined as:

$$E_{\text{dissipation}} = E_{\text{input}} - E_{\text{ATP}}, \quad (16)$$

where $E_{input}$ is the energy released upon translocating 2.67 to 5.67 protons which correspond to 0.53 eV to 1.13 eV. For simplicity, assuming a typical value of 3 protons translocated per molecule of ATP formed would correspond to *ca.* 0.6 eV which can be taken as a working value for the sake of illustration.

The free energy change ($\Delta G$) for ATP synthesis under biological conditions is $\Delta G_{ATP}$ = 30 to 50 kJ/mol.[97,98] Assuming a value of $\Delta G_{ATP}$ = 45 kJ/mol, *i.e.*, around ≈ 0.45 to 0.5 eV, as a typical value, the minimum dissipation (the most efficient operation of the molecular motor) would be $\min(E_{dissipation}) \approx 0.10$ eV and the maximum dissipation (least efficient operation) max $\min(E_{dissipation}) \approx 0.15$ eV. These efficiencies converted to percentages (Dissipation Percentage = $\frac{E_{dissipation}}{E_{input}} \times 100\%$) yield a range of between 17%-25% of the proton's energy is realistically dissipated. Thus, the fifth or quarter of available energy is dissipated through the various channels described above. In this work we have not considered previously overlooked potentially energy-generating effects such as the reinforcement of the proton motive force due to the intrinsic electrostatic potential of the ATP synthase molecule itself.[70]

**Table 1**  Contributions of the principal channels of energy dissipation during the operation of the $F_oF_1$ ATP synthase, with estimated energy loss per proton (in eV) and percent of the total energy available per proton. (Calculated on the basis of $\Delta\psi$ = 200 mV).

| Dissipation Channel | Estimated energy loss per proton | |
| --- | --- | --- |
| | eV | Percent |
| Internal friction ($F_1$ motor) | ~ 0.14 | 70% |
| Proton leak | ~ 0.04 | 20% |
| Maxwell demon dissipation | ~ 0.018 | 9% |
| Structural elasticity/slippage | ~ 0.015 | 7.5% |
| Thermal noise (Brownian motion) | ~ 0.01 | 5% |
| Viscous drag ($F_o$-IMM) | ~ 0.0005 | 0.3% |
| Electroviscous drag | ~ 0.0001 | 0.05% |

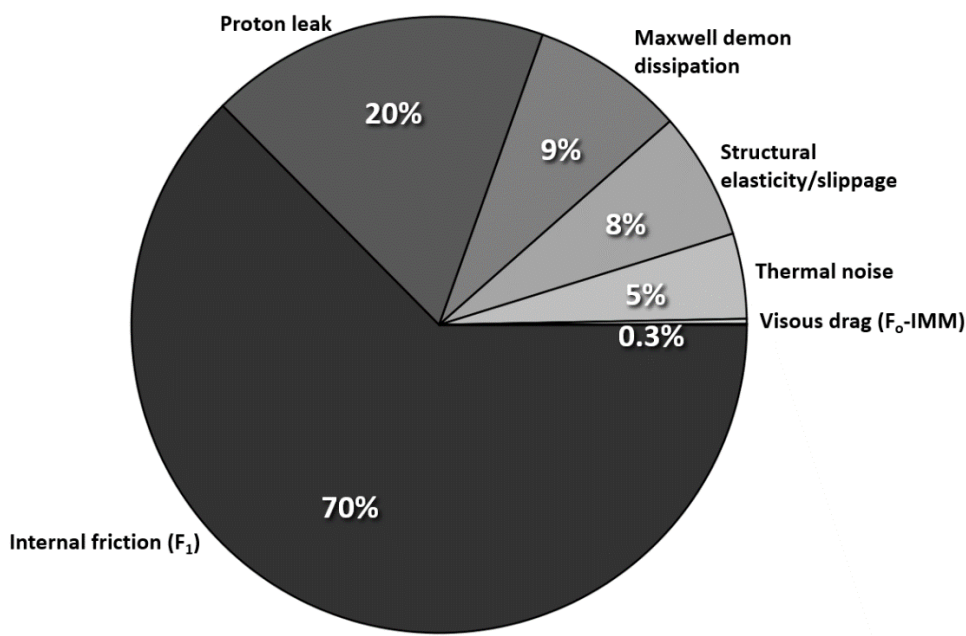

**Fig. 1** Energy dissipation channels in ATP synthase rotation per proton translocated. The dominant dissipation arises from internal mechanical friction within the $F_1$ motor, followed by the proton leak contribution, then the dissipation channel arises from information-theoretic considerations (Maxwell Demon dissipation) reflecting the fundamental energy cost of processing and erasing information during proton translocation, followed by structural elasticity, and random Brownian (thermal) noise, and viscous drag. (Electroviscous drag is too small a sector to be visualized, see Table 1). The values given are the estimated *maximal possible values* for each of these channels, in actual operation of ATP synthase, the values are considerably lower depending on the competition of the various channels and the actual local conditions of operation since some pathways may only operate partially (or not at all).

## 5. Final Thoughts

Our analysis establishes that quantum mechanical constraints, such as angular momentum quantization and tunneling, play no practical role in limiting the rotation of the $F_o$ unit of ATP synthase under biological conditions. Admittedly, modeling the $F_o$ c-ring as a rigid cylinder is a simplifying approximation while in reality ATP synthase exhibits conformational flexibility, internal elastic deformations, and transient coupling to surrounding lipids and protein subunits. However, the moment of inertia primarily depends on mass and radial geometry, both of which remain largely conserved during rotation. While localized or transient deformations could introduce small fluctuations in angular velocity or internal damping, these effects do not materially alter the constraints we examine. Although the rigid rotor model is often introduced for point particles or diatomic molecules, the quantum mechanical treatment applies generally to any rigid body with a well-defined moment of inertia. In this study, we model the $F_o$ c-ring as a hollow cylinder rotating about its central axis, and the standard quantization of angular momentum and rotational energy levels remains valid for this geometry, as long as it satisfies the conditions of rigid body motion.

The calculated energy level spacings are dwarfed by thermal noise, and the tunneling probability through rotational barriers is effectively zero. The $F_o$ unit, rotating at approximately 100–650 rps, is operating far below the quantum mechanical thresholds predicted for a rigid rotor of comparable moment of inertia. Instead, classical dissipative forces dominate, including internal mechanical friction, membrane viscous drag, proton leakage, and even (classical/Shannon/Landauer) information-theoretic energy costs associated with molecular-scale information-erasure/selection processes. Evolution has thus engineered ATP synthase to function robustly within the classical domain, where energy dissipation is minimized but still significant. The enzyme's operation near the thermodynamic efficiency limit of biological energy conversion underscores the sophistication of natural selection at the quantum-classical boundary. Our findings not only clarify the classical nature of ATP synthase but also provide a rigorous framework for the design of future synthetic nanomotors and quantum-classical biomimetic devices. The principal conclusions of this work are captured in Fig. 1 and Fig. 2.

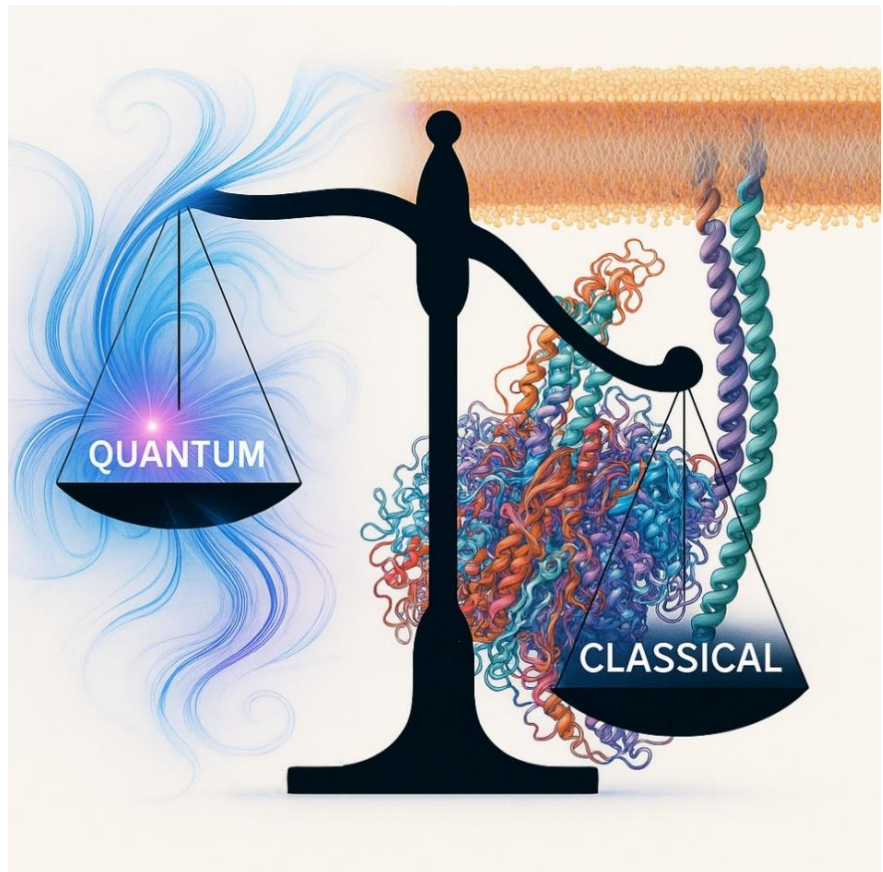

**Fig. 2** ATP synthase as a nanoscale machine at the quantum-classical interface: An artist's depiction of the balance between quantum and classical effects in the operation of mitochondrial ATP synthase. The lighter "quantum" side symbolizes subtle quantum constraints on rotation that are essentially totally smeared and dwarfed by the much "heavier" dominant classical mechanical and thermal processes.

As a final thought, we estimate the critical temperature below which quantum behavior would dominate. From Eqn (7), $\Delta E$ has been estimated to range between approximately $1.8 \times 10^{-10}$ eV and $3.9 \times 10^{-11}$ eV. Using these values and setting $k_B T \approx \Delta E$, we obtain a critical temperature ranging of between $\approx 2.1$ $\mu$K and $\approx 0.5$ $\mu$K, respectively. Such temperatures are unattainable even in cold molecular interstellar clouds which are orders of magnitude warmer (10-20 K), nor even in the coldest known region of interstellar space, the Boomerang Nebula, which has a temperature of about 1 K, which is still approximately 1000 times "hotter" than the critical quantum threshold ot ATP synthase.[99] Consequently, the rotation of the $F_o$ c-ring under physiological or even astrophysical conditions is fully classical. For rotational quantum effects to start playing a role, ATP synthase must be brought to such extremely low temperatures that can only be reached by such techniques as laser cooling[100,101] or Bose-Einstein condensation[102].

The finding that the critical temperature is on the order of 1 microkelvin resonates with Erwin Schrödinger's early insight that living systems must be composed of sufficiently large numbers of atoms to suppress quantum uncertainties and maintain classical stability. As Schrödinger argued in What is Life? (1944), "*only in the cooperation of an enormously large number of atoms do statistical laws begin to control behavior, with quantum fluctuations becoming negligible in large aggregates*". Schrödinger's adage is to be understood as a *special case* (the inner circle of a Venn diagram) of a larger set of effects whereby quantum mechanical effects can manifest themselves despite (or perhaps because of) the cooperation of millions of atoms as occurs in Bose-Einstein condensates of in the magnetic compasses in birds' brain that help orient them in their migratory voyages. This principle in its restricted sense, originally formulated at the dawn of molecular biology, finds direct confirmation in the classical operation of the $F_o$ unit of ATP synthase, exemplifying how life is carefully poised at the quantum-classical boundary.[103]

## Acknowledgements

The authors are also grateful to the Natural Sciences and Engineering Council of Canada (NSERC), the Canadian Foundation for Innovation (CFI), Saint Mary's University, Dalhousie University, Mount Saint Vincent University, Digital Research Alliance of Canada, and Research Nova Scotia for their financial support and resources.